\documentclass[12pt,a4paper,final]{iopart}

\usepackage{amstext}
\usepackage{subfigure}
\usepackage{iopams}  
\usepackage{graphicx}
\usepackage{epstopdf}
\usepackage[breaklinks=true,colorlinks=true,linkcolor=blue,urlcolor=blue,citecolor=blue]{hyperref}

\usepackage{lineno}
\usepackage{threeparttable}
\usepackage{color}
\usepackage[section]{placeins}
\usepackage{comment}
\newcommand{\comm}[1]{}
\usepackage{epstopdf}
\usepackage{float}
\usepackage{multirow}
\usepackage{subfigure}
\usepackage{hyperref}
\usepackage{enumerate}
\modulolinenumbers[5]

\begin{document}
	
\title[\small A Molecular Dynamics Study of Self-Diffusion in Stoichiometric B2-NiAl crystals]{A Molecular Dynamics Study of Self-Diffusion in Stoichiometric B2-NiAl crystals}

\author{Marcin Ma\'zdziarz, Jerzy Rojek, and Szymon Nosewicz}
\address{Institute of Fundamental Technological Research Polish Academy of Sciences,
Pawi\'nskiego 5B, 02-106 Warsaw, Poland}
\ead{mmazdz@ippt.pan.pl}

\begin{abstract}
{Self-diffusion parameters in stoichiometric B2-NiAl solid state crystals were estimated by molecular statics/dynamics simulations with the study of required simulation time to stabilise diffusivity results. An extrapolation procedure to improve the diffusion simulation results was proposed. Calculations of volume diffusivity for the B2 type NiAl in the 1224\,K to 1699\,K temperature range were performed using the embedded atom model potential. The results obtained here are in much better agreement with the experimental results than the theoretical estimates obtained with other methods.}
\end{abstract}

%Uncomment for PACS numbers title message
% Uncomment for keywords
\vspace{2pc}

% Uncomment for Submitted to journal title message
%%\submitto{\MSMSE}
% Comment out if separate title page not required
%\maketitle

\section{Introduction}
\label{sec:Int}

For numerical material modelling, identification of model parameters for the material is one of the key tasks which commonly causes many difficulties. A straightforward approach to parameter determination relies on direct measurement of the wanted parameters.
In some cases, however it is difficult or very expensive to design experiments that allow direct measurements of the required parameters. Alternatively, parameters of the model can be identified by numerical simulations at a lower scale. This methodology used in multiscale modelling is an appealing option, notably when experimental measurements are not feasible or experimental data are not accessed \cite{Cocks-2001}.

This work presents the determination of the diffusion parameters in stoichiometric B2-NiAl crystals using the molecular dynamic simulations.
The nickel-aluminium (NiAl) type intermetallics are modern materials with low density and advantageous mechanical properties.
These are often used as matrix materials in composites manufactured using sintering technologies \cite{NosJoCM}.
There is a growing demand for modelling support of design and optimisation of sintering processes. Sintering is a diffusion driven process, and the knowledge of diffusivity of investigated materials is usually necessary to 
define parameters of sintering models \cite{NosAdvPow}.

Diffusivity can be estimated using both atomic and atomistic modelling.
These approaches can be used to model sintering directly, however, this approach has some limitations.
Direct simulations of particle sintering are limited only to nanoscale, whereas in general, average particle sizes in NiAl powder can be three orders of magnitude bigger, see e.g.\cite{Henz2009,Rojek2017263}. Therefore, atomistic models have been aimed to simulate diffusion in order to evaluate the diffusion properties of NiAl material which can be used in sintering models \cite{NosAdvPow}.
Generally, the diffusivity of NiAl material was derived from the so called, static and dynamic simulations. Using molecular statics (MS) approach and modified analytic embedded-atom method (MAEAM) potential the activation, formation and migration energies of Ni self-diffusion in intermetallic compound NiAl have been calculated for five diffusion mechanisms in \cite{Chen2007102}. However, diffusivity was not analysed during this study. The results show that Ni diffusion is predominated by the triple-defect diffusion mechanism since it needs essentially the lowest migration and activation energy among the five diffusion mechanisms considered.
Point defect energetics and the activation energy in NiAl were determined by molecular statics in \cite{Mishin1997a, Mishin1997b,Mishin1998625,Mishin2003}.
The effective formation energies for a homogeneous thermodynamically stable ordered compound B2-Ni$_x$Al$_{1-x}$ were
determined by a combination of the \textit{ab initio} electron theory with a generalized grand canonical statistical
approach in \cite{Meyer1999, Jiang20052643, Marino20083502}. %  E$_v$=0.74 dla Ni i E$_v$=1.97 dla Al, E$_v$=0.30 dla Ni i E$_v$=1.83 dla Al pamietac ze w obliczeniach DFT mamy formation enthalpy - wzor 6  Musze wiec uzyc dodatkowo energii kohezji dla Al i Ni%
By means of molecular statics simulations and the embedded atom model (EAM) potentials for the assumed six-jump cycle (6JC) mechanism in the temperature spectrum from 800 to 1500\,K \cite{Divinski2000} have found the pre-exponential factor, $D_0$=1.3\,x\,10$^{-5}$(m$^{2}$s$^{-1}$) and the activation energy, $Q$=3.12(eV/atom) ($E_M$=2.44 and $E_F$=0.68).
By means of first-principles density functional theory (DFT) calculations \cite{Marino2008} five postulated diffusion
mechanisms have been analysed for Ni in NiAl in the temperature range from 1200 to 1500\,K and derived the activation energies $Q$=2.99-4.15 (eV/atom) and the pre-exponential factor $D_0$=0.46-1.49\,x\,10$^{-5}$(m$^{2}$s$^{-1}$).
Even for a simpler monatomic fcc Fe system, the DFT calculated self-diffusion $D_0$ is underestimated by two orders of magnitude, see \cite{WANG2018153}. A universal tendency can be seen, that the activation energy $Q$ determined from molecular statics and first-principles are generally in good agreement with experimental data but pre-exponential factor $D_0$ is significantly undervalued,  \comm{2-7 times} see also discussion in Sec.\ref{sec:ADiffNiAl}. 
It is worth noting that such calculations \comm{ordinarily} do not straightforwardly provide the thermal properties of a material \comm{yet}. The harmonic approximation accuracy even taking into account, the influence of temperature on other properties of a material, is far from having reached a satisfactory level \cite{Lejaeghere2014}.
Anharmonic effects can be very significant but there are very few analytic calculations taking into account higher-order terms using perturbation theory,  whereas numerical molecular dynamics calculations can account for anharmonicity to all orders. Some effects associated with thermal expansion at constant pressure can be described by the quasiharmonic approximation but anharmonic effects, which explicitly depend on the magnitude of the atomic vibrational displacements, are present even at fixed volume; see \cite{Forsblom2004}. It seems, therefore, that \comm{only} molecular dynamics is a reliable method that allow one to take into account all anharmonic effects \cite{Grabowski2009}. The experimentally specified Debye temperature of nominally stoichiometric NiAl varies from 470\,K to 560\,K \cite{Lozovoi2003}. In connection to our research, the relevance of the Debye temperature can be twofold. First, it can provide an estimation of the temperature above which all phonon states become occupied. This would mean that below Debye temperature the classical considerations are not likely to apply. Second, the Debye temperature is widely adopted as an estimation of the temperature under which the underestimation of anharmonicity in the QHA, QHAD is not very important, even for systems with reduced symmetry (e.g. with defects).	
The stochastic Monte Carlo method is another static approach to study diffusion. The Monte Carlo simulations represent a broad spectrum of computational algorithms that rely on repeated random sampling to simulate the time evolution of some processes which appear in nature \cite{yip2007handbook}. It is a fairly straightforward and helpful technique that can output a sequence of configurations and the times at which transitions happen between these configurations \cite{weinberger2016multiscale}. Atomistic diffusion in faced-centered cubic NiAl binary alloys, Ni containing 0.05, 0.1, 0.15, 0.2 and 0.25 atomic fraction of Al, was examined by Kinetic Monte Carlo (KMC) method by \cite{Alfonso2015}. Fundamental data obtained from \textit{ab initio} computer calculations  were used as an input to these simulations. The derived activation energies $Q_{Ni}$ were in good agreement with experimental data but pre-exponential factor $D_{0,Ni}$ is unfortunately again highly undervalued.

The most natural dynamic technique to calculate diffusivity is to use molecular dynamic \comm{(MD)}.
An overview of recent molecular dynamics simulations in equiatomic Ni-Al systems can be found in \cite{Evteev2012}. In \cite{Kuhn2014}, diffusion and interdiffusion in binary metallic melts, Al-Ni and Zr-Ni, was analysed by molecular dynamics computer simulations and the mode coupling theory of the glass transition. But the authors claimed that their "\textit{results not to be quantitative predictions for real Al-Ni or Zi-Ni melts, but rather for model systems that allow us to understand the relevant mass transport mechanisms in such melts qualitatively and semiquantitatively}". 
Molecular dynamics study of self-diffusion in liquid Ni$_{50}$Al$_{50}$ alloy was carried out in \cite{Kerrache2008, Levchenko2010331, Levchenko2016} and MD simulation of 2D diffusion in (110) B2-NiAl film in \cite{Evteev2011848}.
The first molecular dynamics \comm{(MD)} simulation of diffusion mechanisms in ordered stoichiometric Ni$_3$Al with the Finnis--Sinclair interatomic potential was performed by \cite{Duan2006}, but the author noted that "\textit{results are at least qualitatively, and in some respect quantitatively, in agreement}".
In a comparative study of embedded-atom methods applied to the reactivity in the Ni–Al system \cite{Turlo2017}, the self-diffusion of the liquid Ni-Al mixture was also studied, unfortunately without giving details of methodology. The consistency of their simulation results with experiment is rather poor. The authors even stated that "\textit{self-diffusion in a solid state is much slower and diffusion coefficients are much smaller by several orders of magnitude than in the liquid state. As a result, MD timescales are generally too short for an extensive study of these phenomena}", what we \comm{deny}  refute in this work. 

The initial estimation of the volume, surface, and grain-boundary diffusivity for the B2 type NiAl as the pivotal mechanism of sintering was performed in \comm{Authors'} our previous work \cite{MazdziarzIJMCE2017}. That paper presented the methodology \comm{of applied method}, the assumption for the model and initial results for three temperatures - from 1573 to 1673\,K. The estimated values were supposed to be applied to discrete element modelling of the sintering process of NiAl powder, however such small range of temperatures can be insufficient and can lead to inconsistency of material data in the context of the connection between atomistic and microscopic scales. In response to such challenges, the current work concentrates on improved and extended investigations of volume diffusion determination in NiAl material in the much wider temperature range - from 1224 to 1699\,K. Determined materials parameters will be characterized \comm{by} to better accuracy and agreement with experimental results. Furthermore, the present paper discusses the inspection of the necessary simulation time required to ensure the stabilization of diffusivity results. The studied problem seems to be the one of the major issue of proper determination of diffusivity results and have a considerable impact on obtained final results. Therefore, it has been proposed the new procedure to improve the quality of diffusion simulation results, what is important especially in the analysis of diffusion results in lower temperatures.

\section{The mechanism of atomistic diffusion in NiAl}
\label{sec:ADiffNiAl}

\begin{figure}[!htb]
	\centering
	\includegraphics[width=.38\linewidth]{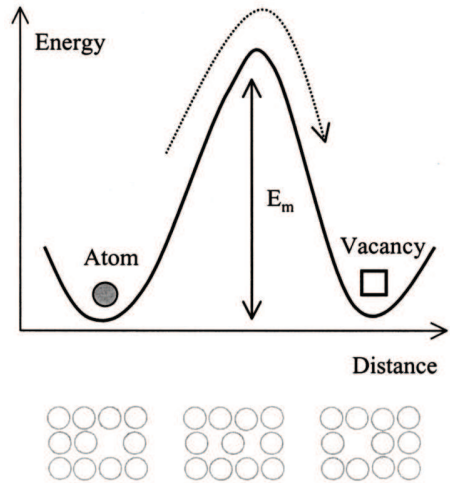}
	\caption{Schematic representation of the volume self-diffusion process.}
	\label{fig:Diffusion}
\end{figure}

Diffusion is material transport induced by the motion of atoms. A schematic illustration of volume self-diffusion in the monatomic structure of an atom, from its initial position into a vacant lattice site, is depicted in Fig.\ref{fig:Diffusion}. The migration energy, $E_M$, has to be applied to the atom in order to \comm{so that it could} overcome inter-atomic bonds and to move to the new position.

Similarly to other solids, self-diffusion is the \comm{most} primary diffusion process in intermetallics \comm{as well}.
In binary compounds two tracer self-diffusion coefficients -- one for A atoms and \comm{another} one for B atoms, -- are relevant \cite{mehrer2007diffusion}. In the NiAl binary system, we can identify Ni self-diffusion and Al self-diffusion, where the first one is highly dominative in the stoichiometric composition \cite{gupta2010diffusion, Herzig2004993, Ramunni2014112}.
Interdiffusion, or chemical diffusion, is different from the tracer diffusion, because it is present in a
chemical composition gradient and not in a homogeneous solid \cite{mehrer2007diffusion}.

For the B2 type NiAl, see Fig.\ref{fig:Samples}a), various hop mechanisms based on experimental and theoretical considerations have been suggested \cite{Xu2010}. In addition to the dominant nearest-neighbour (NN) hops \comm{dominating} in monoatomic structures, next-nearest neighbour (NNN) hops are also feasible in B2-NiAl due to its open crystal structure. An intriguing property of B2-NiAl is that NN Al hops from the Al sublattice to the Ni sublattice cannot emerge, as the end state is predicted \comm{not} to be not mechanically stable. Other suggested mechanisms include two simultaneous pair-atom (SPA) hop mechanisms, the six-jump cycle (6JC), two anti-structure bridge (ASB) mechanisms and the triple-defect (TD) sequence.

It is well-known \comm{widely known} that \comm{the} thermal defects occur at  finite temperatures apart from the constitutional defects.
Modelling concepts are  usually based on the idea of a noninteracting point defects gas as suggested by Wagner and
Schottky. Thermal defects in an ordered binary alloy of a fixed composition must occur in a balanced manner in order to retain the alloy stoichiometry.
Hence,  since the alloy composition is fixed, they can occur merely in the composition-conserving combination and any \comm{of the} point defects cannot be a thermal defect alone. There are four basic composition-conserving defects composed of two types of point defects: \textit{Exchange antisite defect} (X), \textit{Divacancy defect} (D), \textit{Triple Ni or simply triple defect} (TN) and \textit{Triple Al defect} (TA), see \cite{Korzhavyi2000}. It is worth mentioning that the formation energies of single point defects depend on the choice of the reference states and they are not insignificant, as opposed to the formation energies of composition-conserving defects, which do not depend on a specific choice of the reference state and can be directly compared. Thereby, the first case requires a grand canonical ensemble, the second a canonical ensemble. In the B2-type intermetallics the thermal disorder is typical of a triple defect type and the triple defect (TN) formed by one antisite Ni atom and two Ni vacancies is presumed to be the dominant thermal defect in NiAl, see \cite{Korzhavyi2000, Herzig2004993}.

The temperature dependence of diffusivity typically follows an Arrhenius equation \cite{mehrer2007diffusion} and typically is written as:
\begin{eqnarray}
{D}={D}_{0}\,\exp\left(-{\frac{Q}{k_BT}}\right) ={D}_{0}\,\exp\left(-{\frac{E_F+E_M}{k_BT}}\right) ,
\label{eqn:Arrhenius}
\end{eqnarray}
or in the logarithmic form
\begin{eqnarray}
{\log}\left({D}\right) ={\log}\left({D}_{0}\right)-\left({\frac{Q}{k_B}}\right)\left({\frac{1}{T}}\right) ,
\label{eqn:ArrheniusLog}
\end{eqnarray}
where  $D$ is the diffusivity or diffusion coefficient; $D_{0}$ is the pre-exponential factor; $Q$ is the activation energy; $T$ is the temperature; $k_B$ is the Boltzmann's constant; $E_F$ is the formation energy and $E_M$ is the migration energy.

The thermal equilibrium concentration of point defects is given by the relation: 
\begin{eqnarray}
{\frac{N_D}{N}} = \exp\left(-{\frac{E_F}{k_BT}}\right) ,
\label{eqn:Ecpd}
\end{eqnarray}
where  $N_D$ is the number of defects and $N$ is the number of potential defect sites.

Simple Arrhenius behaviour should not however be deemed to be universal. At  $T>$ 1500\,K  an upward deviation from this equation can be noted, among others, in B2-NiAl \cite{ Divinski2000, Korzhavyi2000} and bcc-Fe \cite{Mendelev2009}.

The activation energy $Q$ for the stoichiometric composition for triple Ni defect (TN) in the Arrhenius Eq.\ref{eqn:Arrhenius} leads to:
\begin{eqnarray}
{Q}=\left({\frac{E_{TNi}}{3}}\right)+E_M ,
\label{eqn:Acten}
\end{eqnarray}
where  $E_{TNi}/{3}$= $E_{F}$, see \cite{gupta2010diffusion}.

Presently, there exist no directly measured Al tracer diffusion data in NiAl, Al diffusion cannot be determined experimentally because of the absence of a suitable isotope, but Ni diffusion in single-crystal NiAl has been determined \cite{Frank20011399}. For Ni in B2-NiAl self-diffusion prefactor $D_0$=2.71-3.45\,x\,10$^{-5}$(m$^{2}$s$^{-1}$) and activation energy $Q$=2.97-3.01(eV/atom). It is visible that the spread of results is quite marginal as opposed to results of self-diffusion in pure Al and in pure Ni. The diffusion prefactor $D_0$ for Al varies from three orders of magnitude up to eight orders for Ni in different studies \cite{Campbell20115194}.

\section{Computational methodology}
\label{sec:Cm}

The molecular simulations in this study were made with the use of the Large-scale Atomic/Molecular Massively Parallel Simulator (LAMMPS) \cite{Plimpton1995} and visualized in the Open Visualization Tool (OVITO) \cite{Stuk2010} and Wolfram Mathematica \cite{Mathematica}.

%\subsection{Molecular statics simulations}
\subsection{Molecular potentials}
\label{sec:Mp}

Two embedded-atom method (EAM) potentials: EAM2002\cite{Mishin2002} and EAM2009\cite{Purja2009} have been validated for their possible reproduction of NiAl diffusion in the temperature spectrum from 1224 to 1699\,K. Using molecular statics approach  \cite{Tadmor2011, MYD2010, Maz2011, Cholewinski2014, Mazdziarz2015} the following parameters
of NiAl were probed:  lattice constant, cohesive energy, bulk modulus, elastic constants, surface energy, defect energies, migration energies as well as melting point temperature, see Tab.\ref{tab:EAMmp}.
For MS simulations the sample was assumed to be cube and composed of 10$^3$ elementary B2-NiAl cells, see Fig.\ref{fig:Samples}a) ($\sim$ 2000 atoms). The size of the sample allows to fulfil the assumed by Wagner and Schottky condition of non-interaction of point defects.  Energy minimization with periodic boundary conditions (PBC) applied to all faces of the sample was carried out  with \comm{the Polak-Ribiere version of} a nonlinear conjugate gradient algorithm \cite{Plimpton1995}. The assumption was, that convergence was reached when the relative change in the energy and forces, between two successive iterations was less than 10$^{-13}$. Following the energy minimization procedure an external pressure was applied to the simulation box to see the volume change effect.

\subsection{Diffusivity simulations}
\label{sec:Ds}

Periodic boundary conditions (PBC) were applied in all simulations. After molecular statics relaxation, molecular dynamics equilibration for the time period of 1\,ns (0.5x10$^6$ MD steps, one step = 2\,fs) was done. The data gathering simulation covered a time interval of minimum 1000\,ns (500x10$^6$ MD steps). Finding the needed simulation time warranting the stabilization of the results was the first step in the simulation.
At a given temperature and pressure, NPT (constant number of atoms, pressure and temperature) Nose-Hoover style barostat was used \cite{Plimpton1995, Tadmor2011}. The diffusivity \comm{mean-squared displacement (MSD)} was calculated as the average:
\begin{eqnarray}
{D_{avg.}} \equiv {\frac{1}{2d}}\lim_{t\to\infty} {\frac{\left\langle \left[ r(t_0+t)-r(t_0)\right]^2  \right\rangle }{t}} ,
\label{eqn:DiffAv}
\end{eqnarray}
or the instantaneous one:
\begin{eqnarray}
{D_{inst.}} \equiv {\frac{1}{2d}}\lim_{t\to\infty} {\frac{\partial\left\langle \left[ r(t_0+t)-r(t_0)\right]^2  \right\rangle }{\partial t}} ,
\label{eqn:DiffInst}
\end{eqnarray}
where, $d$ is the dimensionality ($d$=2 for surface and grain-boundary diffusion, $d$=3 for volume); $t$ is the time and $\left\langle \left[ r(t_0+t)-r(t_0)\right]^2  \right\rangle$ is the ensemble average MSD. Thermodynamic informations were calculated and stored at intervals of 2\,ps (1000 MD steps).

\subsubsection{Volume diffusivity}
\label{sec:Vds}

\begin{figure}[!htb]
	\centering
	\includegraphics[width=.48\linewidth]{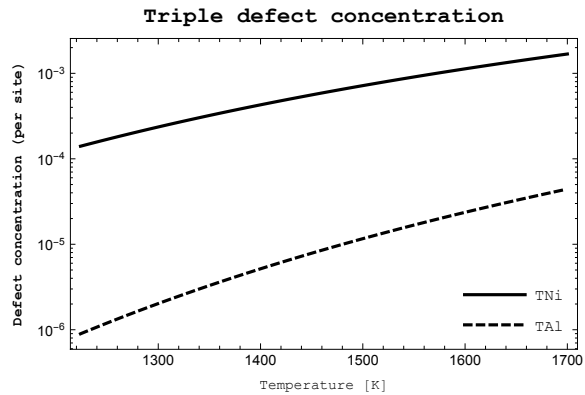}
	\caption{Triple defect concentration for Ni and Al.}
	\label{fig:TDC}
\end{figure}

Thermal equilibrium point defect concentration is a function of temperature and the formation energy $E_F$, see Eq.\ref{eqn:Ecpd}. Defect concentration for triple Ni and Al defect energy found by the molecular potential applied in the MD studies, see Tab.\ref{tab:EAMmp}, in the temperature interval from 1224 to 1699\,K is depicted in Fig.\ref{fig:TDC}.

\begin{figure}[!htb]
	\centering
	\begin{tabular}{cc}
		\includegraphics[width=0.50\linewidth]{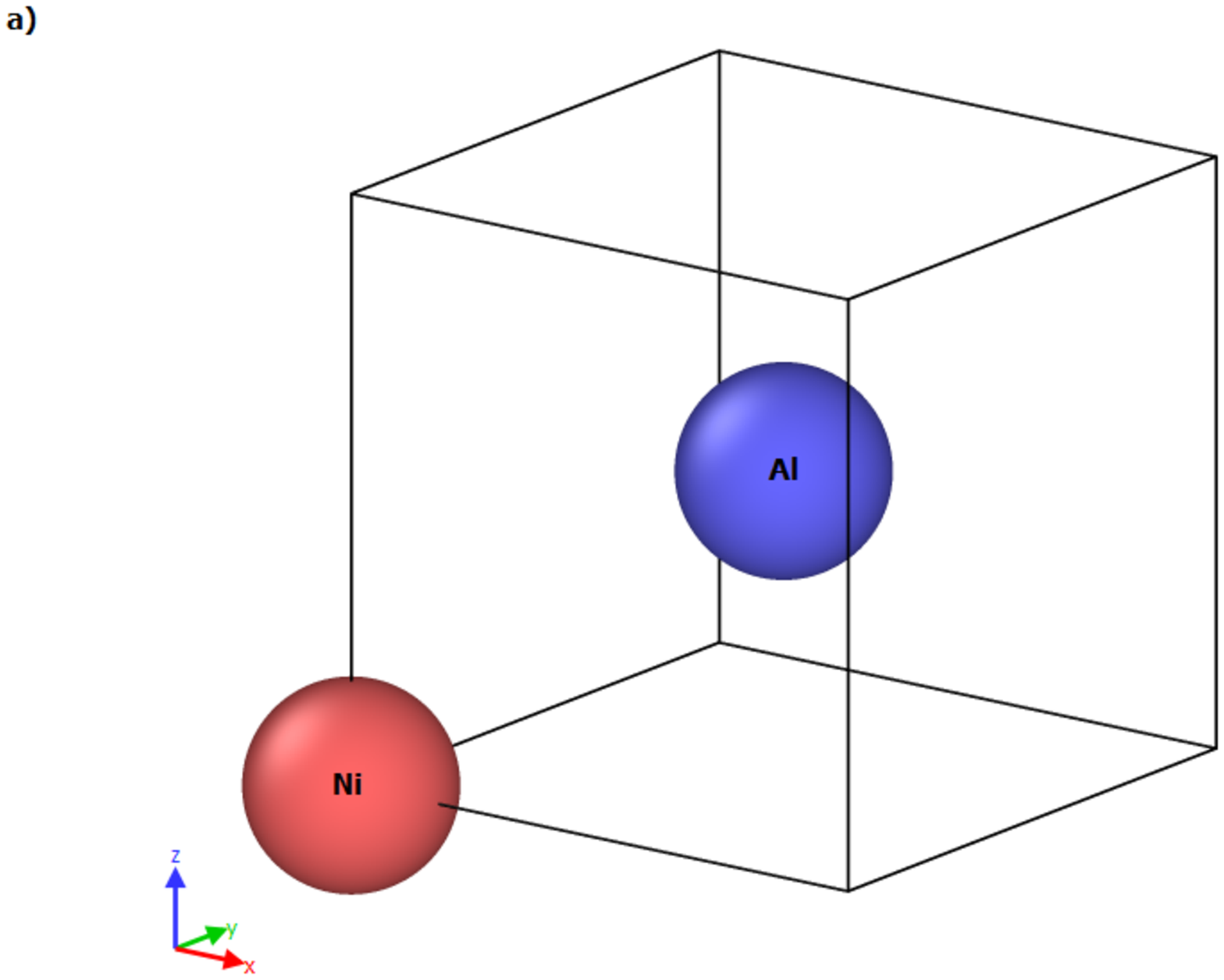} &
		\includegraphics[width=0.50\linewidth]{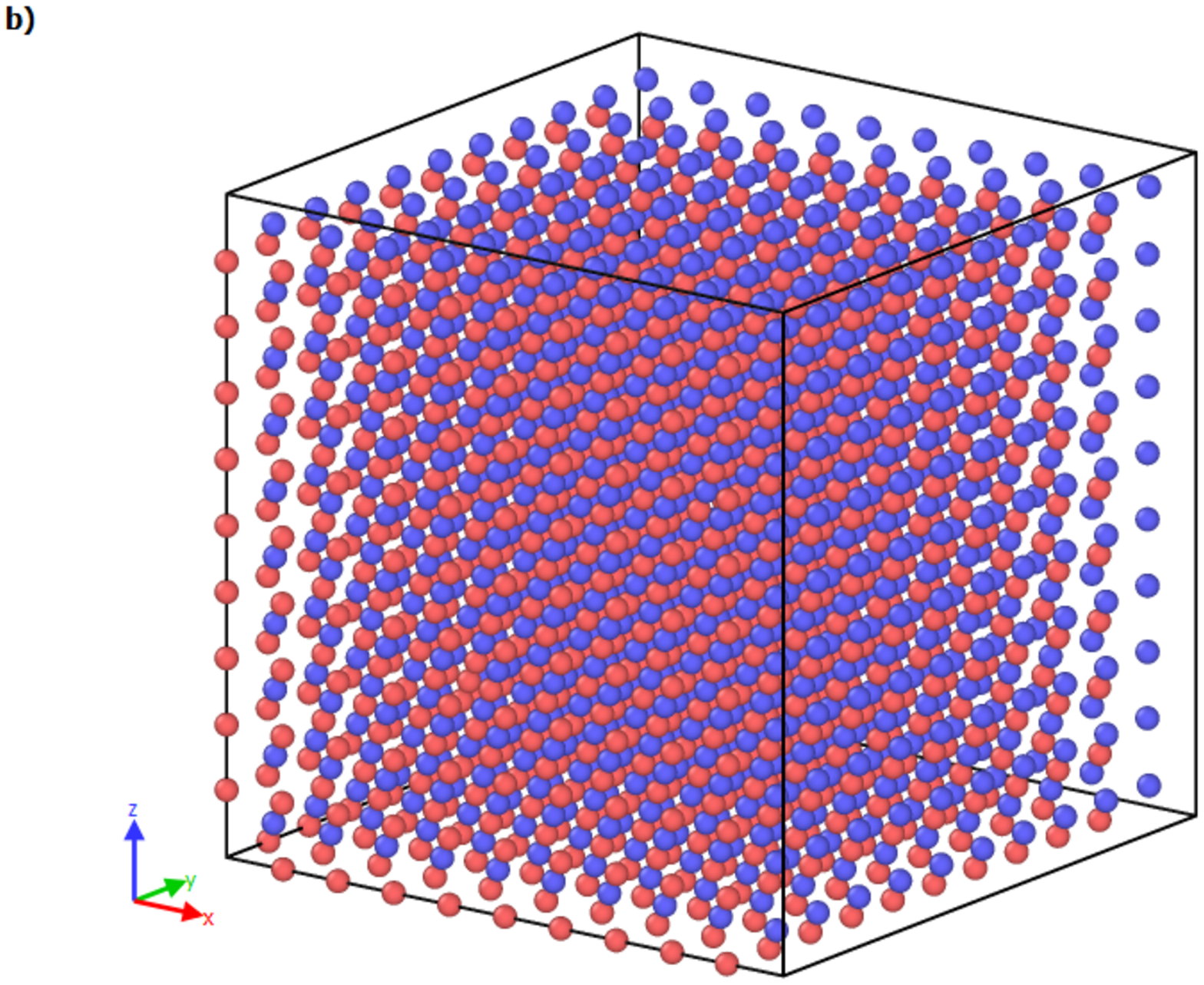}
	\end{tabular}
	\caption{Atomistic models: a) NiAl basic cell, b) simulation box for volume diffusion.  }
	\label{fig:Samples}
\end{figure}

Hence, for the cubic computational sample, Fig.\ref{fig:Samples}b), using above Eq.\ref{eqn:Ecpd} we can calculate that one triple Ni defect (TN) for temperature 1224\,K is in the $22^3$ basic cells (21296 atoms), for 1244\,K $21^3$ basic cells (18522
atoms), for 1265\,K $20^3$ basic cells (16000 atoms), for 1288\,K $19^3$ basic cells (13718 atoms), for 1313\,K $18^3$ basic cells (11664 atoms), for 1341\,K $17^3$ basic cells (9826 atoms), for 1372\,K $16^3$ basic cells (8192 atoms), for 1406\,K $15^3$ basic cells (6750 atoms), for 1445\,K $14^3$ basic cells (5488 atoms), for 1489\,K $13^3$ basic cells (4394 atoms), for 1540\,K $12^3$ basic cells (3456 atoms), for 1599\,K $11^3$ basic cells (2662 atoms) and for 1699\,K $10^3$ basic cells (2000 atoms), respectively. It is worth emphasising that the triple defect concentration for Al is about 100 times lower than for Ni, see Fig.\ref{fig:TDC}, what would mean the use of about 100 times larger computational samples to calculate Al self-diffusion.  

\subsection{Diffusivity determination from simulation results}
\label{sec:Ddfsr}

In order to answer the question of what value of diffusion $D$ should be read from the simulation, two procedures were proposed and tested:

\begin{enumerate}[I]%for capital roman numbers.
	\item \label{itm:first} Direct: we take the value of the diffusivity $D$, Eqs.\ref{eqn:DiffAv},\ref{eqn:DiffInst}, from the end of the MD simulation.
	\item \label{itm:second} Extrapolation: we approximate the results of the diffusivity $D$, Eqs.\ref{eqn:DiffAv},\ref{eqn:DiffInst}, versus simulation time $t$ with a function with a limit, e.g. $D(t)$=$a$$\times$$\exp$($b$/$t$), and take its value for ${t\to\infty}$. 
	%	\medskip
\end{enumerate}

\section{Results}
\label{sec:Res}

Both considered EAM interatomic potentials , see Tab.\ref{tab:EAMmp}, similarly reproduce the parameters of B2-NiAl,
nevertheless, EAM2002 potential predicts the melting point at 1520\,K since the experimental value is 1911\,K. Because of our interest in the temperature spectrum from 1224 to 1699\,K for subsequent molecular dynamics computations the EAM2009 potential, predicting a more reasonable melting temperature of 1780\,K, was used.

\begin{table}
	\caption{\label{tab:EAMmp}Material parameters of NiAl from two EAM potentials - Molecular statics simulations. DFT and experimental data taken from \cite{Mishin2002,Purja2009}. Migration energy$^*$ calculated for nearest-neighbour (NN) hop mechanism.}
	%	\begin{indented}
	\begin{tabular}{@{}llll}
		\br
		{} & EAM2002\tnote{a} & EAM2009\tnote{b} & Experiment/DFT\\
		\mr
		Lattice constant NiAl  [\AA] & 2.86 & 2.832 & 2.88 \\
		Cohesive energy [eV] & -4.47 & -4.51 & -4.50 \\
		Bulk modulus B [GPa] & 160 & 159 & 158 \\
		Elastic constant C$_{11}$ [GPa] & 200 & 191 & 199 \\
		Elastic constant C$_{12}$ [GPa] & 140 & 143 & 137 \\
		Elastic constant C$_{44}$ [GPa] & 120 & 121 & 116 \\
		Surface energy [J$m^{-2}$] & 1.52 & 2.07 & $\sim$2.2 \\	
		Exchange energy $E_X$ [eV] & 2.765 & 2.075 & 2.65-3.15 \\
		Divacancy energy  $E_D$ [eV] & 2.396 & 2.569 & 2.18-3.07 \\
		Triple Ni defect energy $E_{TNi}$ [eV] & 2.281 & 2.807 & 1.58-2.83 \\
		Triple Al defect energy $E_{TAl}$ [eV] & 5.276 & 4.406 & 5.44-6.46 \\
		Migration energy$^*$ $E_M^{Ni}$ [eV] & 2.324 & 2.49 & 2.58 \\
		Migration energy$^*$ $E_M^{Al}$ [eV] & 1.476 & 1.76 & $\sim$1.6 \\	
		Melting point [K] & 1520 & 1780 & 1911 \\
		\br
	\end{tabular}
	%\\
	%\textsuperscript{\emph{a}} Ref. \cite{Mishin2002,Purja2009}
	%	\end{indented}
\end{table}

There is no general rule for how long a simulation has to be run before we reach the long-time asymptotic behaviour predicted by the Einstein-Smoluchowski relation, Eqs.\ref{eqn:DiffAv},\ref{eqn:DiffInst}.
As far as we know the required simulation duration ensuring the stabilization of the slope of the mean-squared displacement (MSD) results was not duly discussed in the literature. For the molecular dynamics analysis of self-diffusion in bcc Fe \cite{Mendelev2009}, the simulation time was $\sim$ 41\,ns, for hcp and bcc Zr \cite{Mendelev2010} $\sim$ 20-30\,ns and for ordered stoichiometric Ni$_3$Al in \cite{Duan2006} was $\sim$ 30\,ns.
For the molecular dynamics simulation of carbon diffusion in $\alpha$-iron, the simulation time was $\sim$50\,ns. Similarly, as we can observe in this work, diffusivity decreases with increasing time and moves towards stabilisation, see \cite{TAPASA20071}.

The conditions where molecular dynamics simulations can be used to calculate highly converged Arrhenius plots for substitutional alloys with various vacancy concentrations between 1\,\% and 5\,\% were analysed in \cite{ZHOU2017331}. It was found that highly converged results can be obtained when using an elevated temperature range (i.e. $T_{simulation}$/$T_{melting}$=0.87-0.98) and an extended simulation time longer than 300\,ns. The higher temperature in the simulations, as well as the higher concentration of defects, resulted in more convergent results and lower statistical error.

\begin{figure}[H] 	
	\centering
	\begin{tabular}{cc}
		\includegraphics[width=0.46\linewidth]{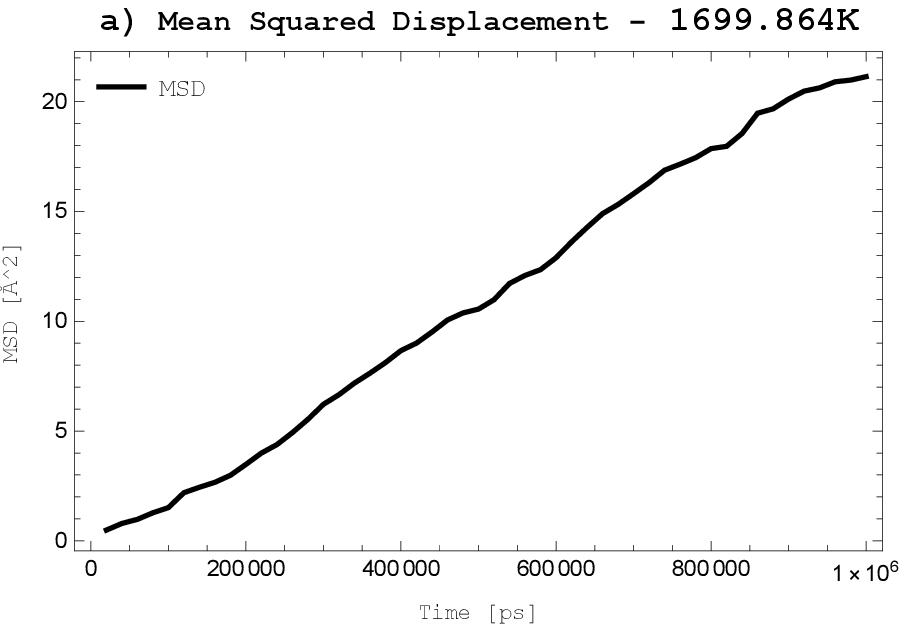} &		
		\includegraphics[width=0.50\linewidth]{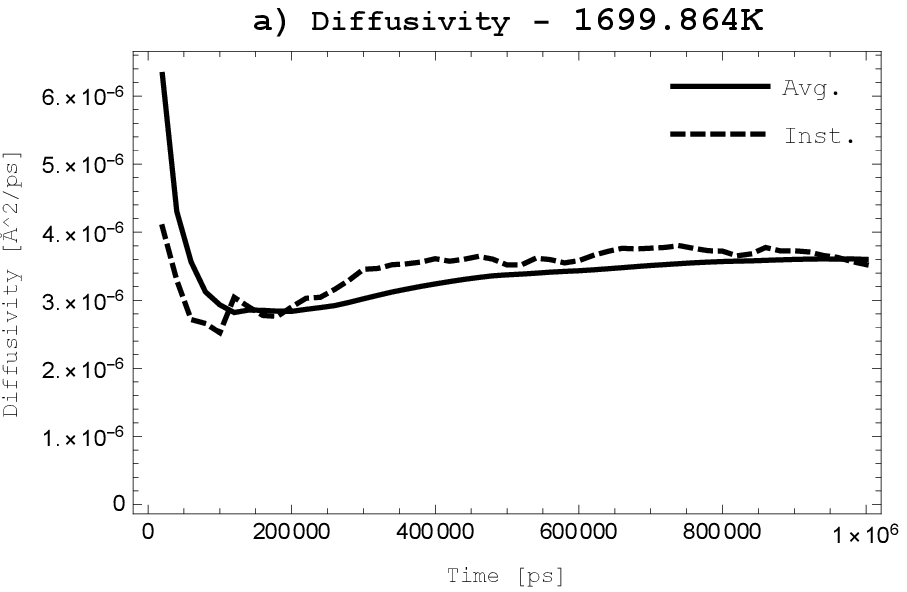} \\

		\includegraphics[width=0.46\linewidth]{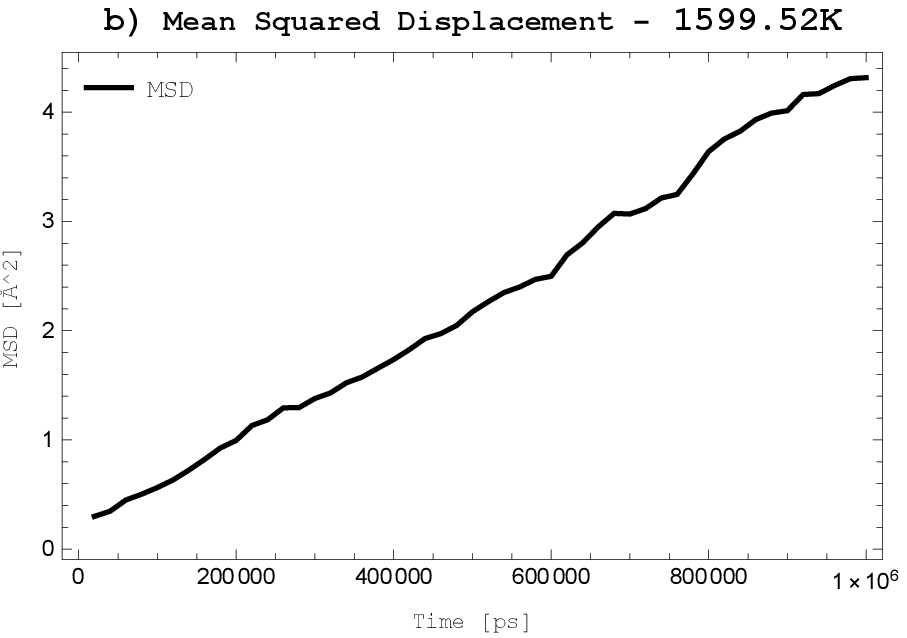} &
		\includegraphics[width=0.50\linewidth]{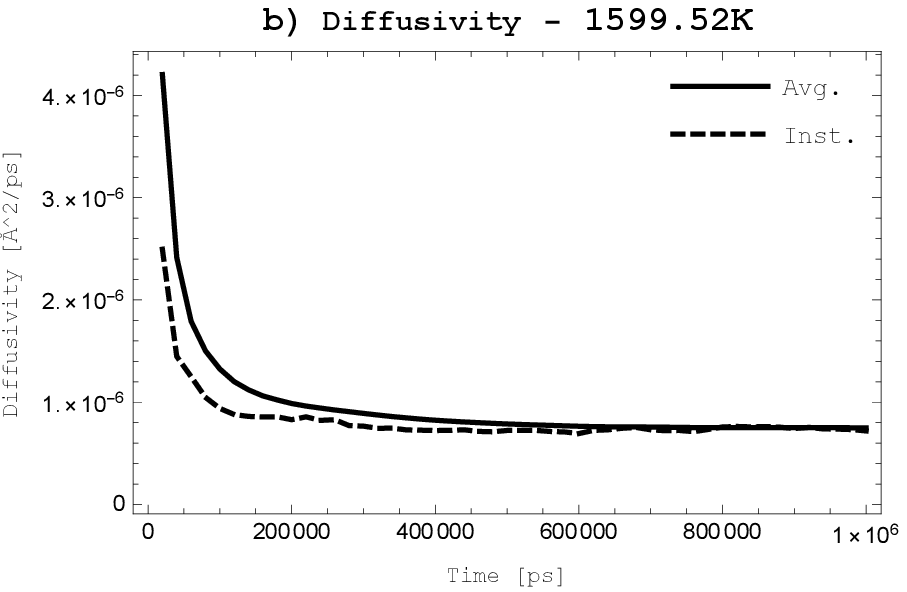} \\

		\includegraphics[width=0.46\linewidth]{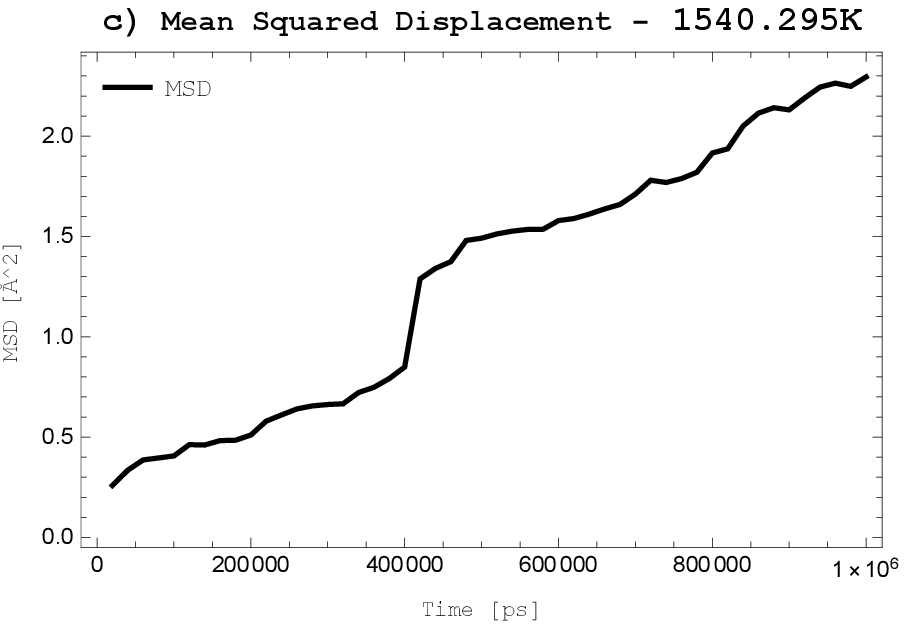} &
		\includegraphics[width=0.50\linewidth]{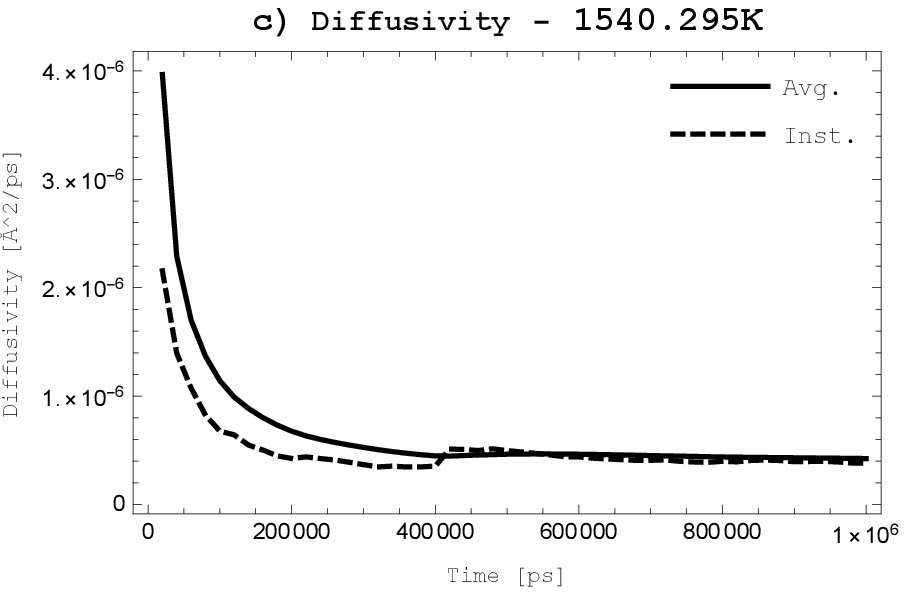} \\

		\includegraphics[width=0.46\linewidth]{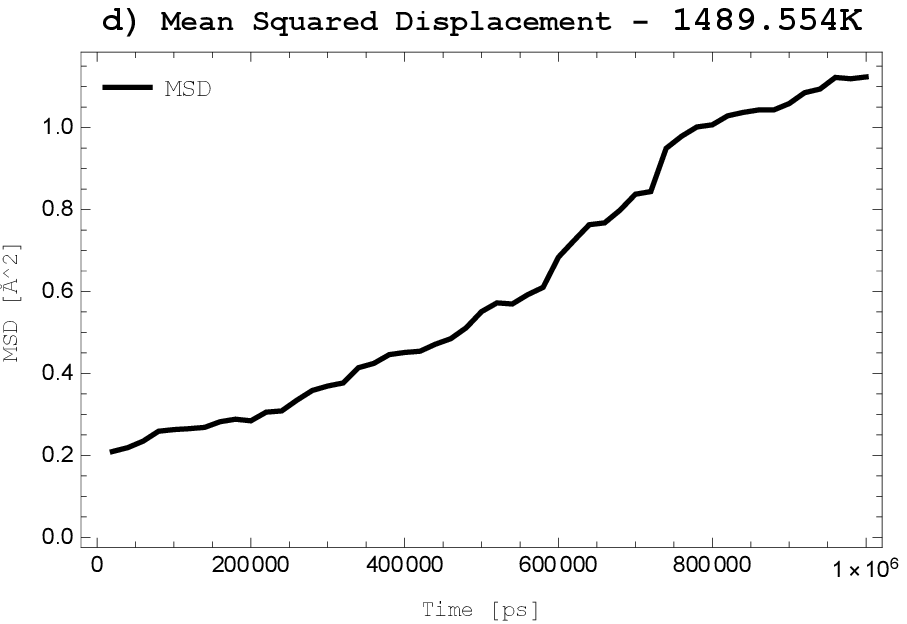} &
		\includegraphics[width=0.50\linewidth]{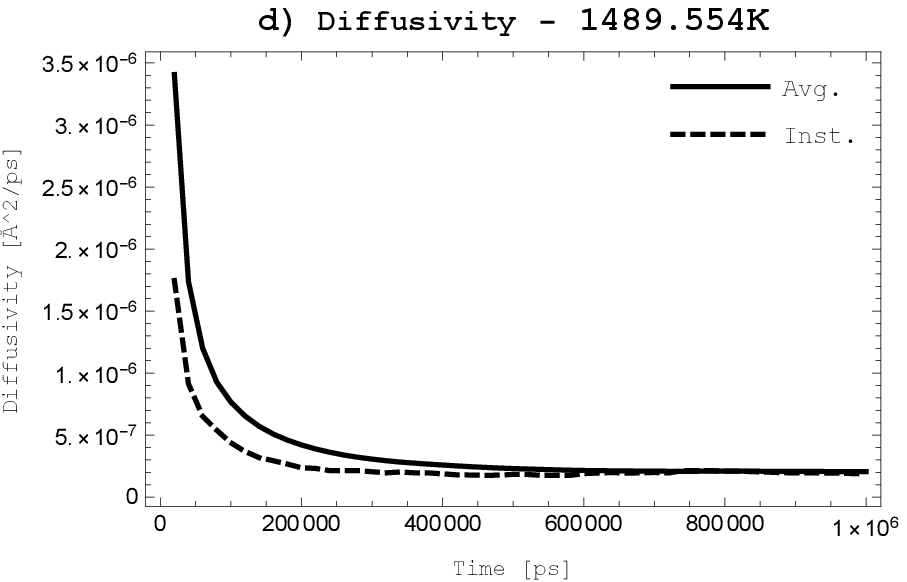} 
	\end{tabular}
	\caption{Mean Squared Displacement and Average and Instantaneous Diffusivity for different temperature (1699-1489\,K).}
	\label{fig:MSDDiffusivity1}
\end{figure}

\begin{figure}[H] 	
	\centering
	\begin{tabular}{cc}
		\includegraphics[width=0.46\linewidth]{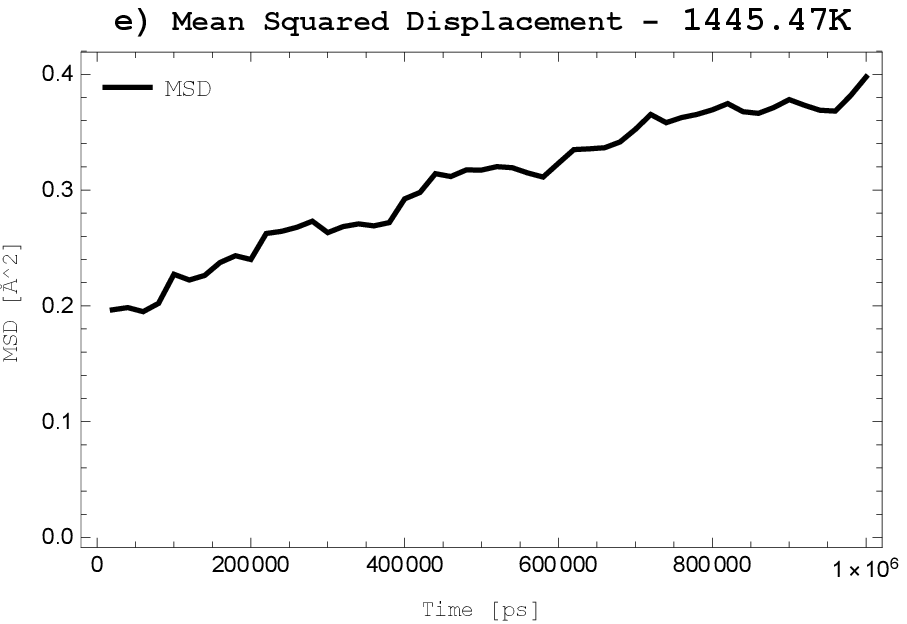} &
		\includegraphics[width=0.50\linewidth]{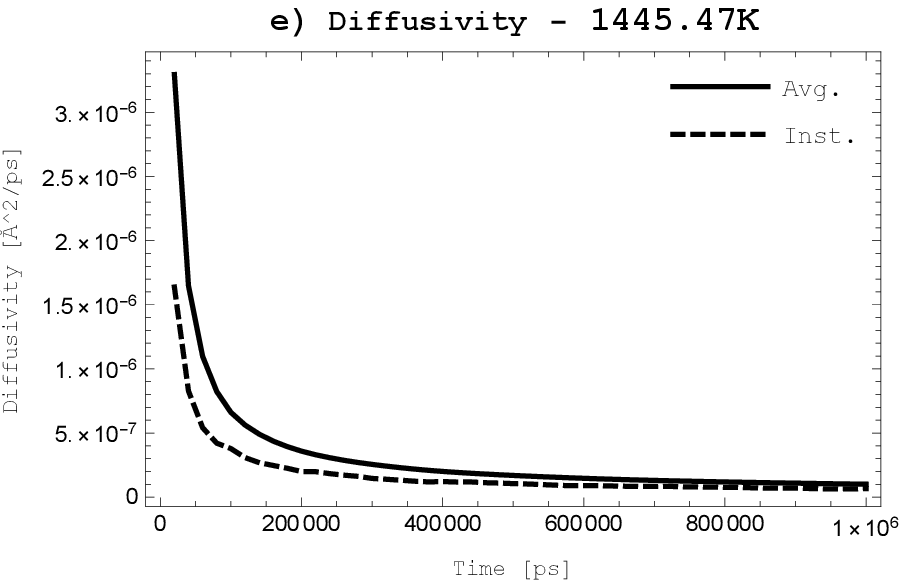} \\

		\includegraphics[width=0.46\linewidth]{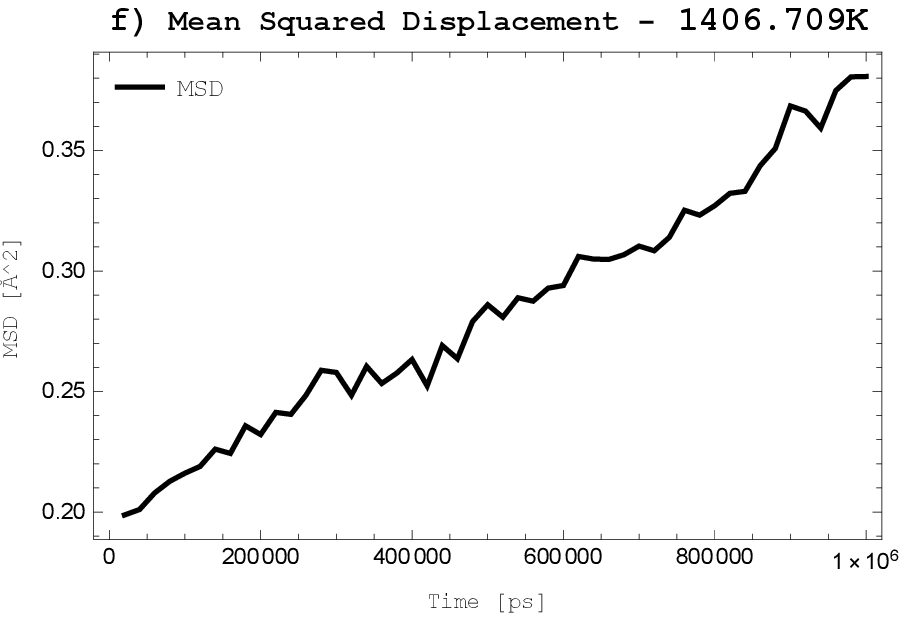} &
		\includegraphics[width=0.50\linewidth]{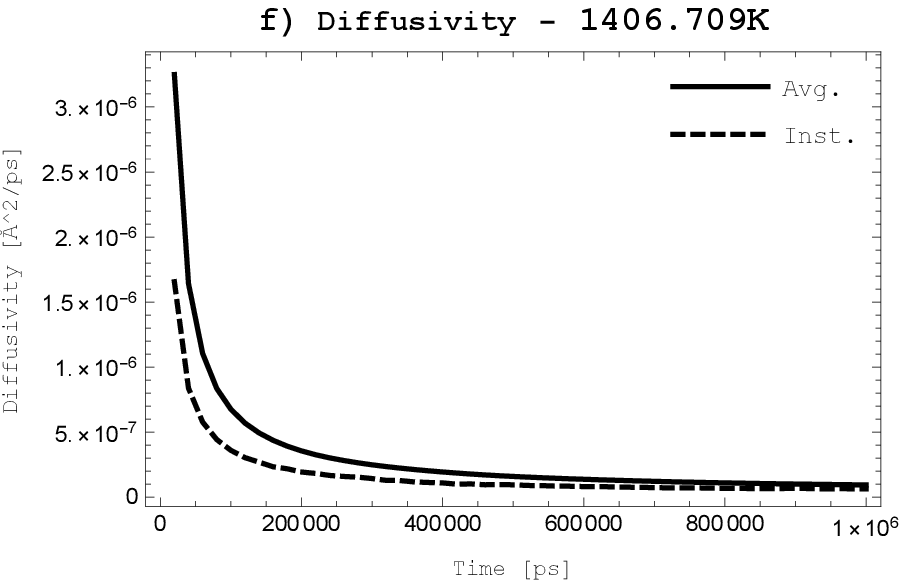} \\

		\includegraphics[width=0.46\linewidth]{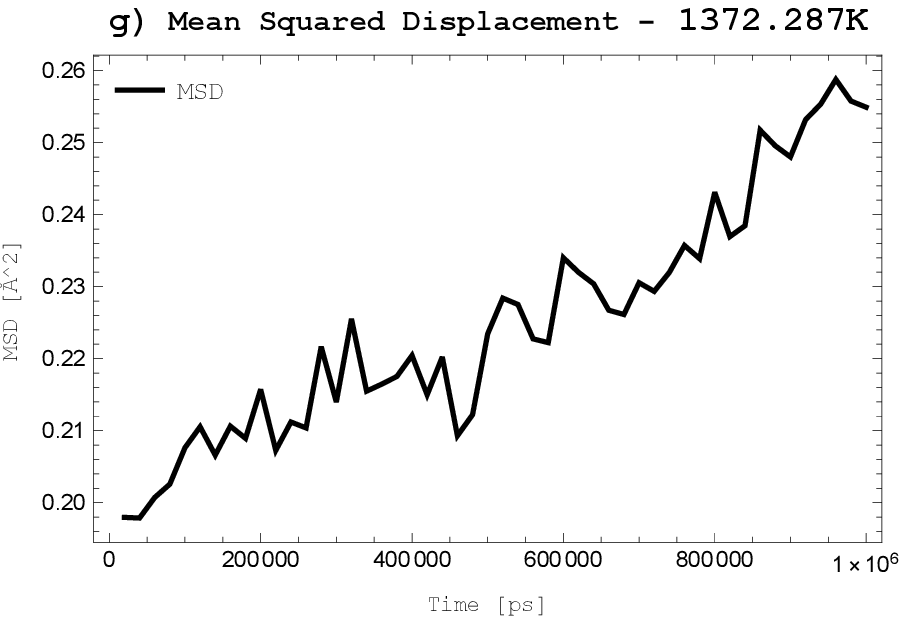} &
		\includegraphics[width=0.50\linewidth]{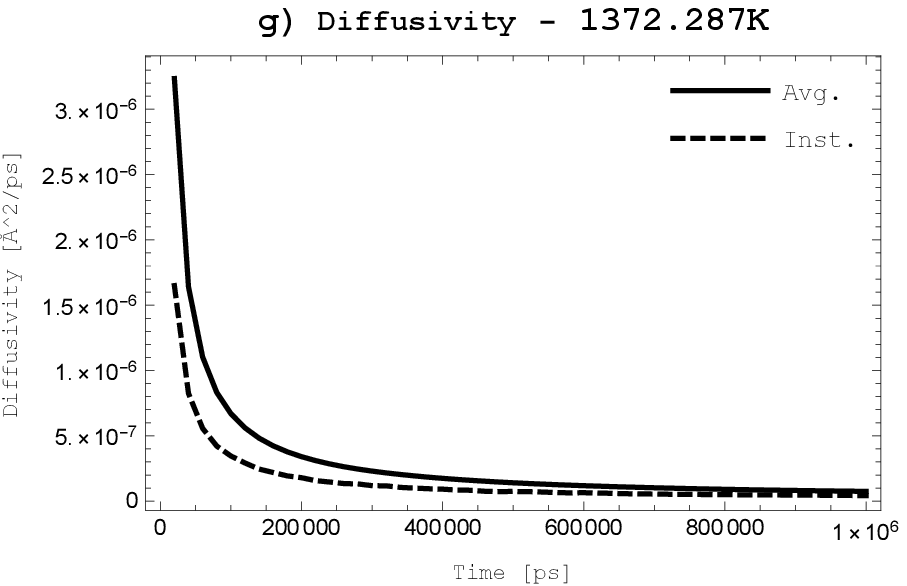} \\

		\includegraphics[width=0.46\linewidth]{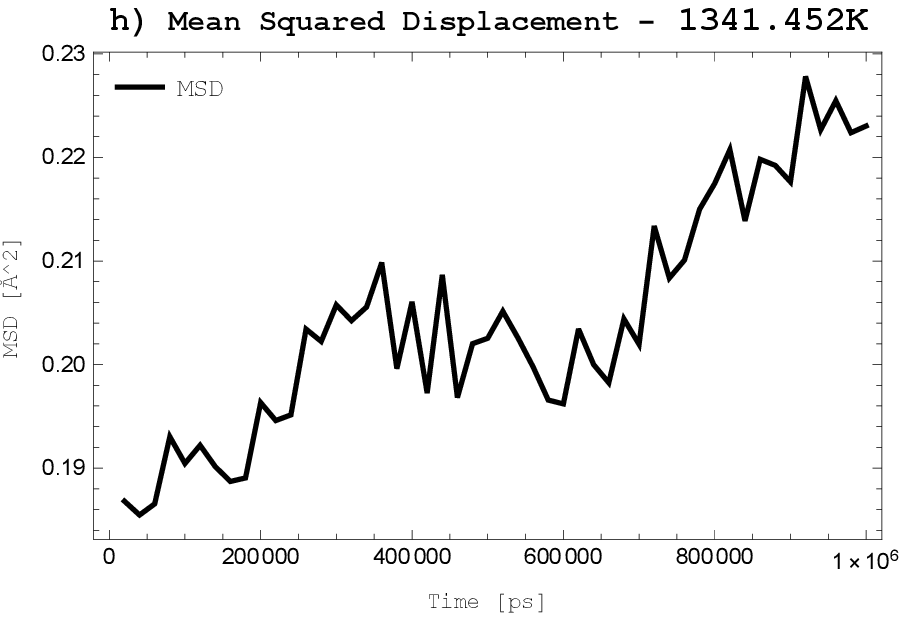} &
		\includegraphics[width=0.50\linewidth]{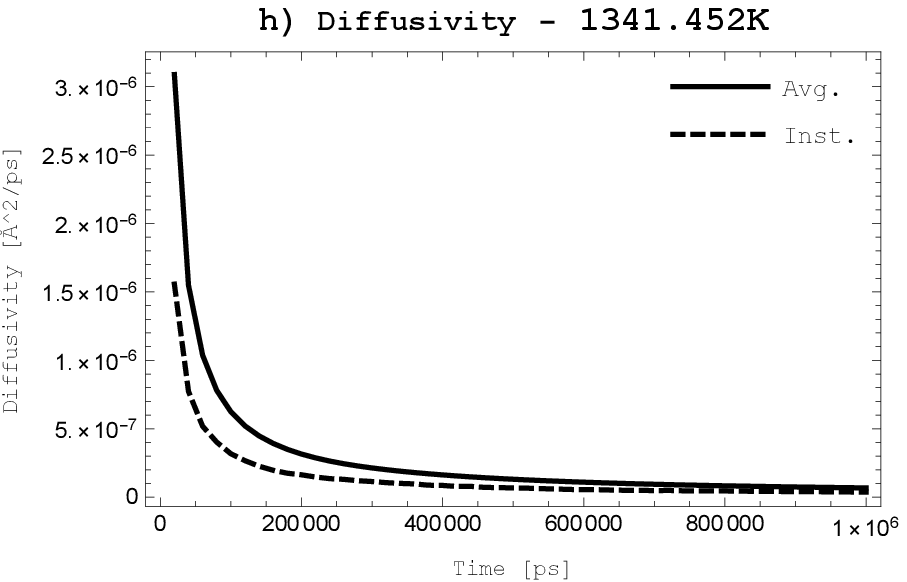} 
	\end{tabular}
	\caption{Mean Squared Displacement and Average and Instantaneous Diffusivity for different temperature (1445-1341\,K).}
	\label{fig:MSDDiffusivity2}
\end{figure}

\begin{figure}[H] 	
	\centering
	\begin{tabular}{cc}
		\includegraphics[width=0.46\linewidth]{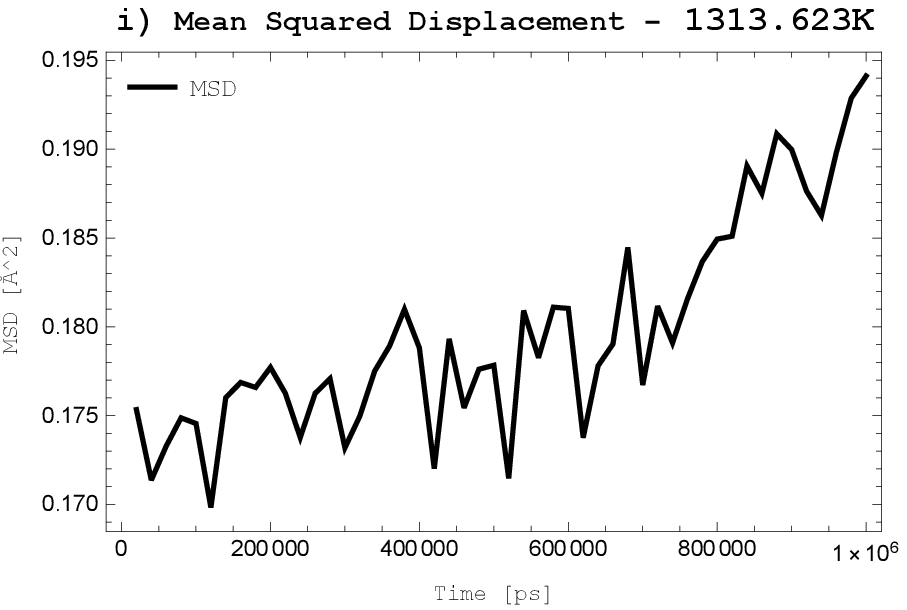} &
		\includegraphics[width=0.50\linewidth]{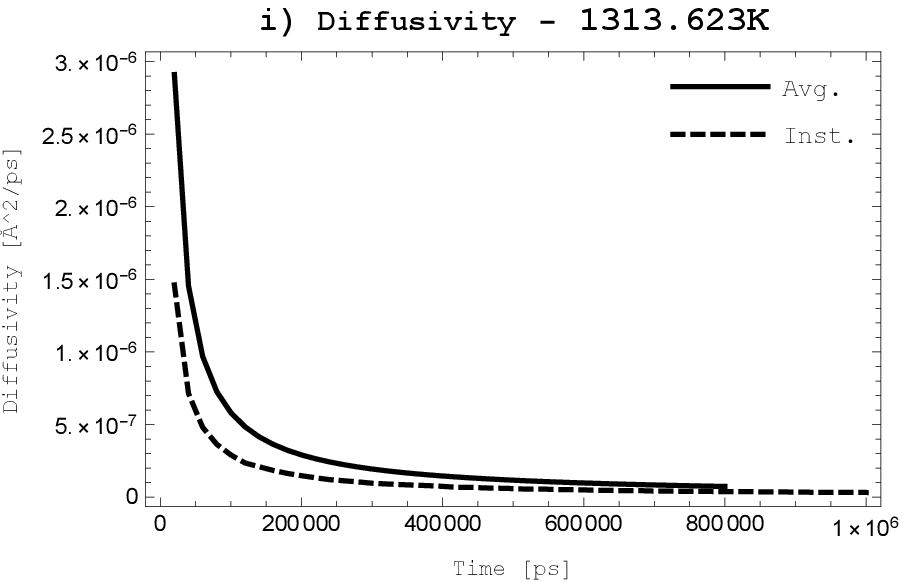} \\

		\includegraphics[width=0.46\linewidth]{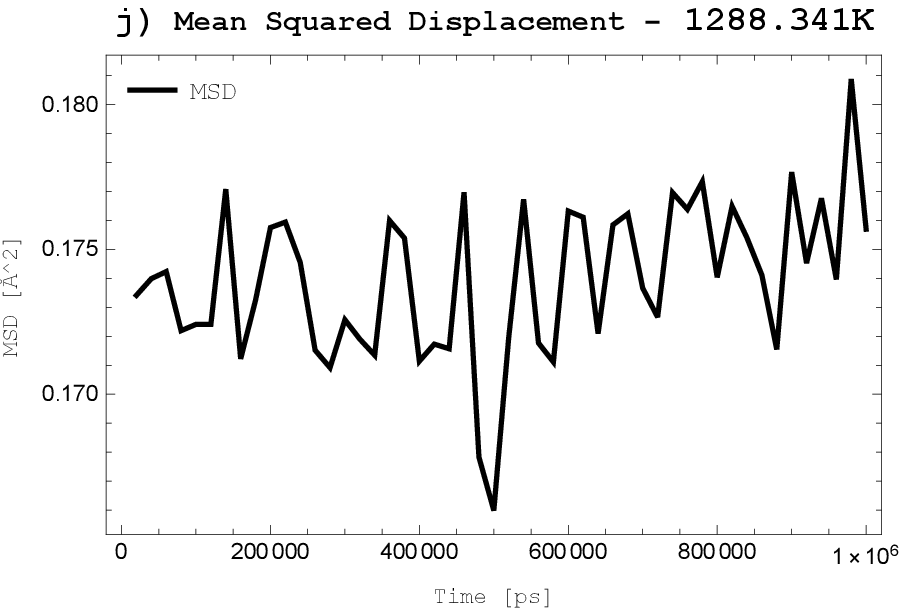} &
		\includegraphics[width=0.50\linewidth]{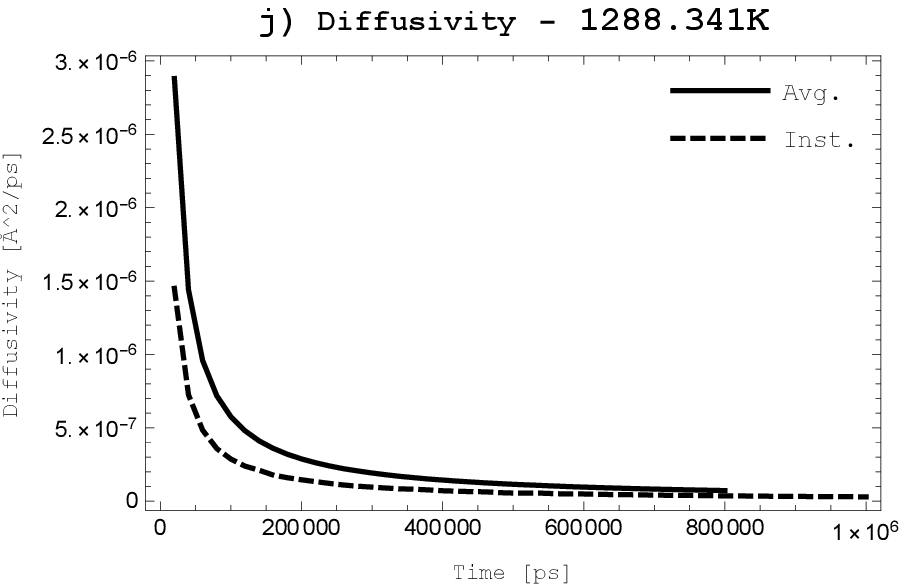} \\

		\includegraphics[width=0.46\linewidth]{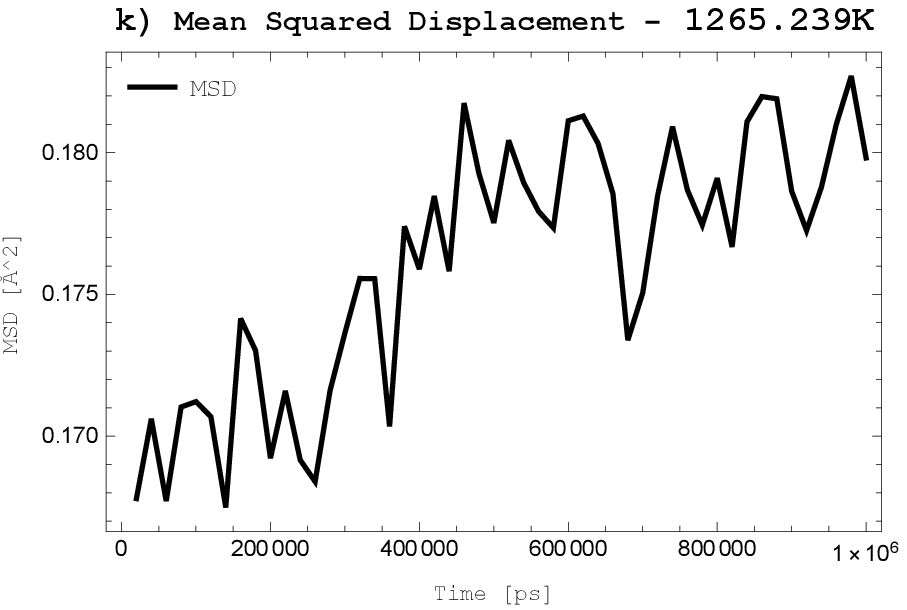} &
		\includegraphics[width=0.50\linewidth]{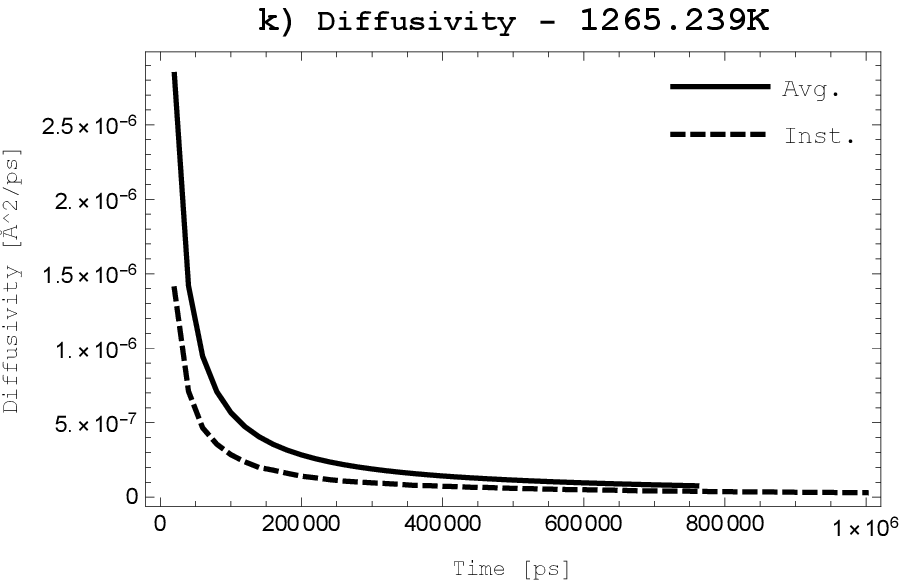}
	\end{tabular}
	\caption{Mean Squared Displacement and Average and Instantaneous Diffusivity for different temperature (1313-1265\,K).}
	\label{fig:MSDDiffusivity3}
\end{figure}

\begin{figure}[H] 	
	\centering
	\begin{tabular}{cc}
		\includegraphics[width=0.46\linewidth]{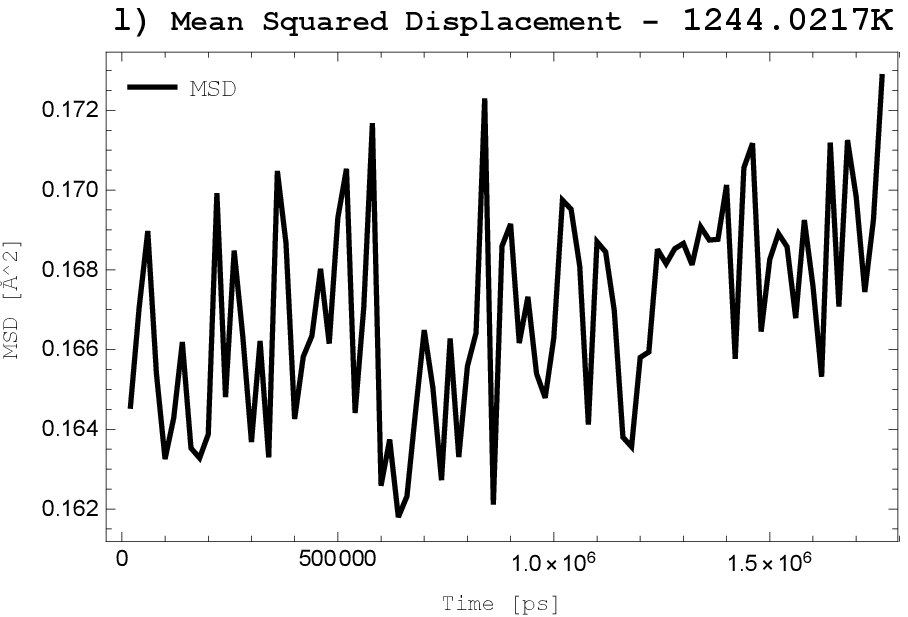} &
		\includegraphics[width=0.50\linewidth]{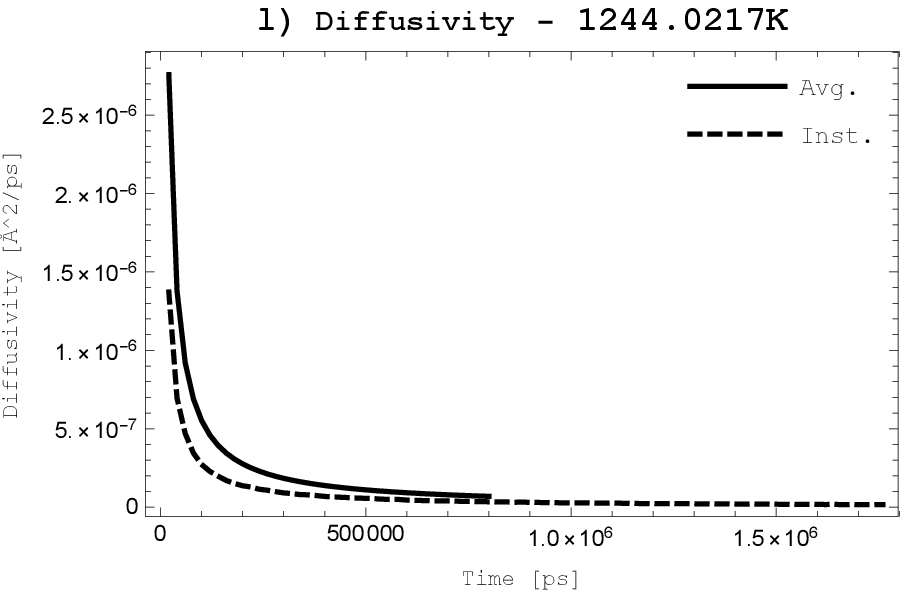} \\

		\includegraphics[width=0.46\linewidth]{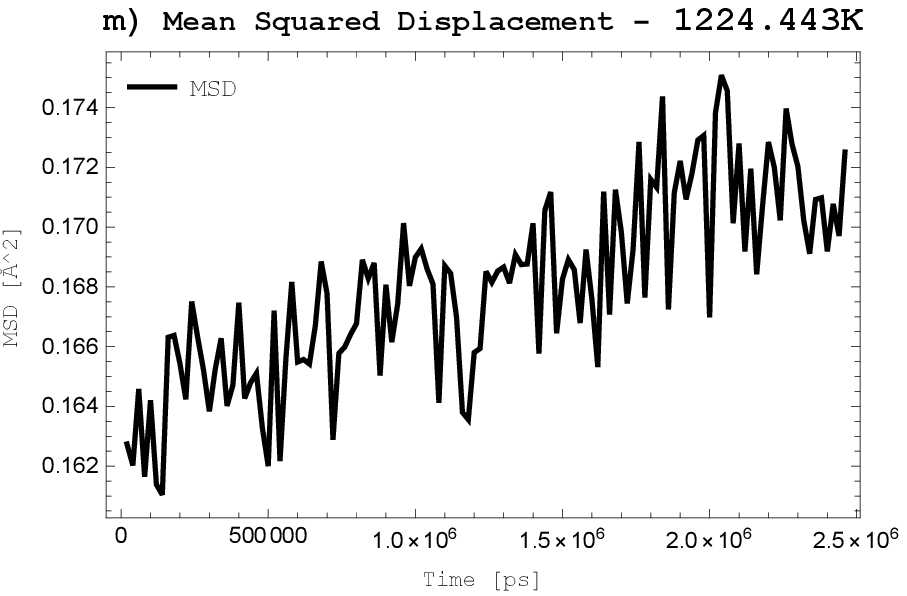} &
		\includegraphics[width=0.50\linewidth]{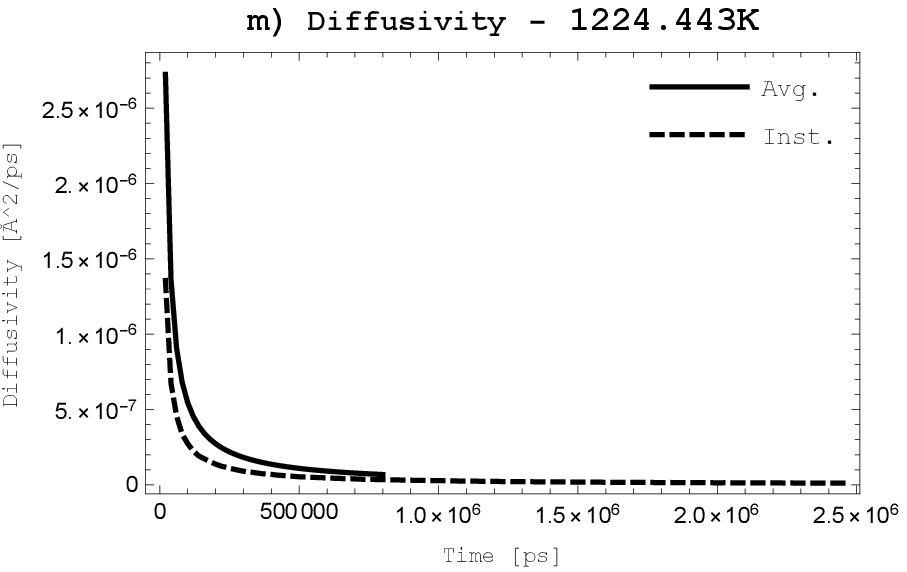}
	\end{tabular}
	\caption{Mean Squared Displacement and Average and Instantaneous Diffusivity for different temperature (1244-1224\,K).}
	\label{fig:MSDDiffusivity4}
\end{figure}

Our observations suggest that the needed simulation time that allows the stabilization of the MSD slope is substantially longer than that used by mentioned authors. It can be further observed in \comm{We can further see in} Figs.\ref{fig:MSDDiffusivity1}-\ref{fig:MSDDiffusivity4}, that the needed simulation time depends on temperature and as it grows, the time decreases. We also observe, that instantaneous diffusivity, Eq.\ref{eqn:DiffInst}, stabilizes faster than the average diffusivity, Eq.\ref{eqn:DiffAv}, and will be used further as a reference value. It is also observed that in conducted calculations the simulation time at least 1000\,ns is needed. For temperature 1244\,K the simulation time had to be extended to 1800\,ns and to 2500\,ns for 1224\,K, respectively.

\begin{figure}[!htb]
	\centering
	\begin{tabular}{cc}
		\includegraphics[width=0.50\linewidth]{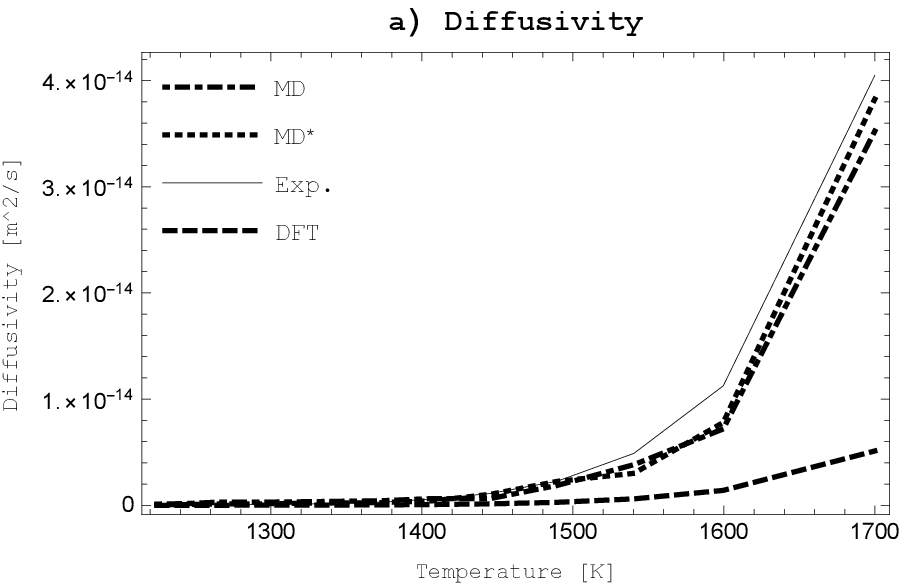} &
		\includegraphics[width=0.48\linewidth]{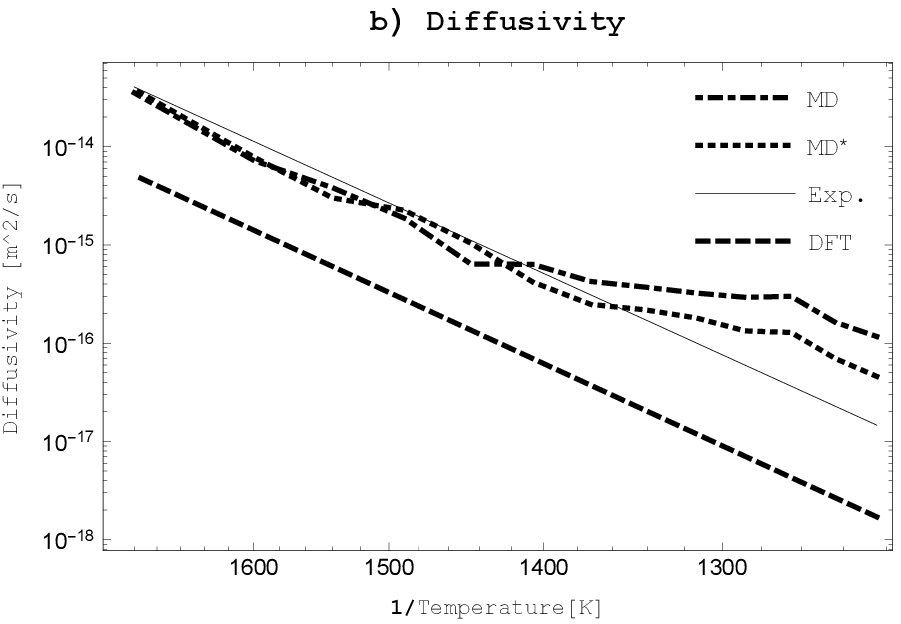}
	\end{tabular}
	\caption{Volume diffusivity for different temperatures: a) normal scale, b) Arrhenius plot. }
	\label{fig:DiffusivityTemp}
\end{figure}

The volume diffusivity for different temperatures from our MD computations, using Direct(\ref{itm:first}) and Extrapolation(\ref{itm:second})  procedure of diffusivity determination from MD simulation results, is summarized in Tab.\ref{tab:DiffTemp} and Fig.\ref{fig:DiffusivityTemp}. Our results fit better to the \comm{are much better in accordance with the} averaged experimental data, \cite{Frank20011399}, than these from DFT calculations for triple defect mechanism (TN) \cite{Marino2008}. The conformity of our simulation results to experimental results from 1700\,K to $\sim$1400\,K is excellent. The use of the Extrapolation procedure(\ref{itm:second}) improves diffusion simulation results especially for lower temperatures. At lower temperatures, the agreement is worse, \comm{unfortunately it is getting worse} but better still than the results of the DFT calculations.

\begin{table}
	\caption{\label{tab:DiffTemp}Volume diffusivity for different temperatures from our molecular dynamics computations (MD$^{\text{I}}$, Direct procedure \ref{itm:first} and MD$^{\text{II}}$, Extrapolation procedure \ref{itm:second}), DFT \cite{Marino2008} calculations and from the experiments \cite{Frank20011399}.}
	%	\begin{indented}
	\begin{tabular}{@{}lllll}
		\br
		Temperature[K] & MD$^{\text{I}}$-$D$[m$^2$/s] & MD$^{\text{II}}$-$D$[m$^2$/s] & Exp.-$D$[m$^2$/s] & DFT-$D$[m$^2$/s] \\
		\mr
		1224 & 1.1688E-16 & 4.5937E-17 & 1.4659E-17	 & 1.7081E-18 \\
		1244 & 1.6368E-16 & 7.001E-17  & 2.2895E-17	 & 2.6801E-18 \\
		1265 & 2.9969E-16 & 1.2856E-16 & 3.6542E-17	 & 4.2985E-18 \\
		1288 & 2.9300E-16 & 1.3336E-16 & 5.9743E-17	 & 7.0636E-18 \\
		1313 & 3.2353E-16 & 1.8037E-16 & 1.0030E-16	 & 1.1923E-17 \\
		1341 & 3.7173E-16 & 2.1841E-16 & 1.7347E-16  & 2.0738E-17 \\
		1372 & 4.2486E-16 & 2.4826E-16 & 3.1014E-16  & 3.7299E-17 \\
		1406 & 6.3452E-16 & 4.1901E-16 & 5.7570E-16  & 6.9678E-17 \\
		1445 & 6.3920E-16 & 1.0412E-15 & 1.1153E-15  & 1.3590E-16 \\
		1489 & 1.8723E-15 & 2.255E-15  & 2.2690E-15  & 2.7852E-16 \\
		1540 & 3.8250E-15 & 3.0116E-15 & 4.8868E-15  & 6.0459E-16 \\
		1599 & 7.1946E-15 & 7.7964E-15 & 1.1251E-14  & 1.4039E-15 \\
		1699 & 3.5237E-14 & 3.8241E-15 & 4.0480E-14  & 5.1172E-15 \\
		\br
	\end{tabular}
	%	\end{indented}
\end{table}

\begin{figure}[!htb]
	\centering
	\begin{tabular}{cc}
		\includegraphics[width=0.48\linewidth]{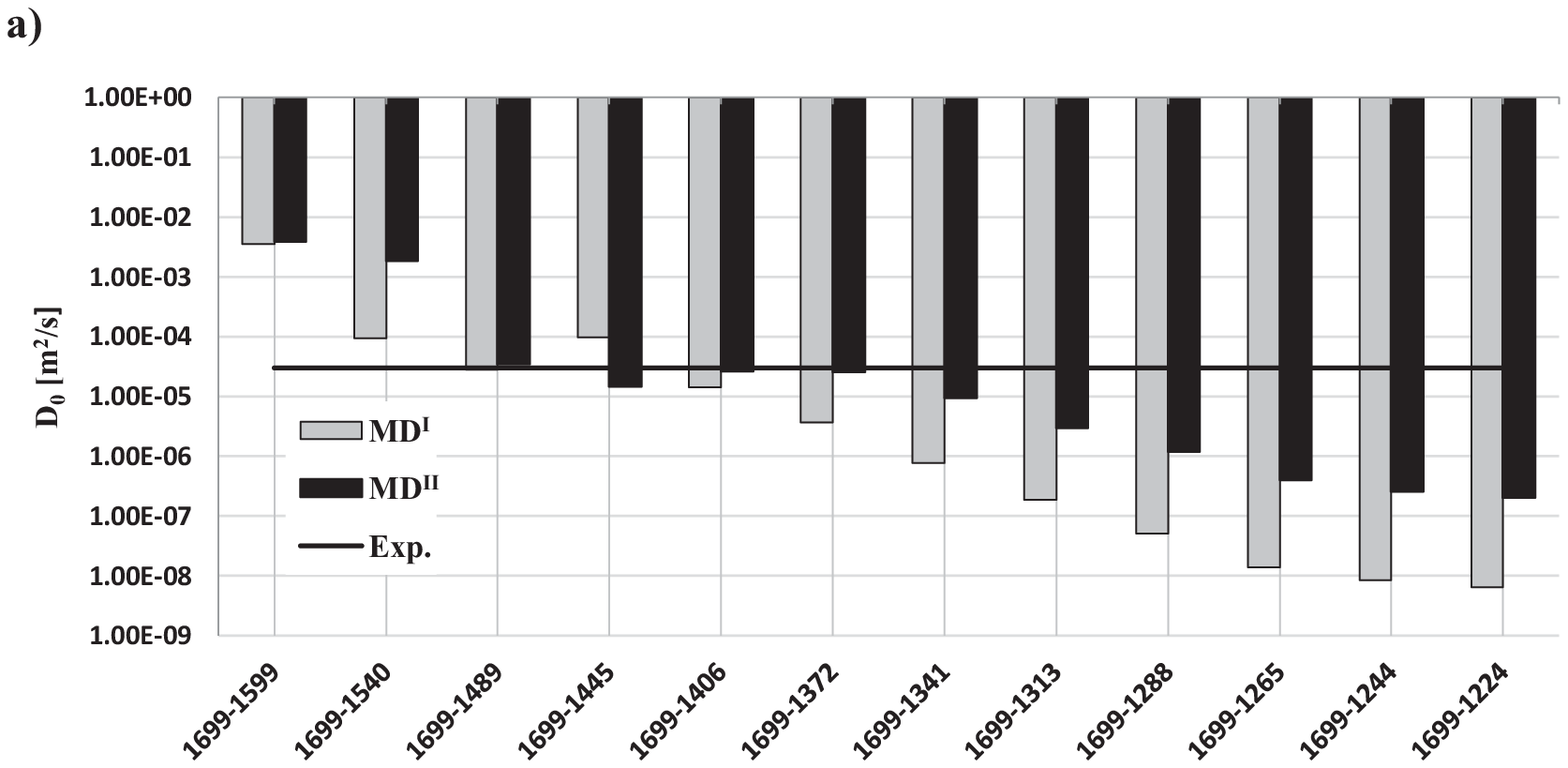} &
		\includegraphics[width=0.48\linewidth]{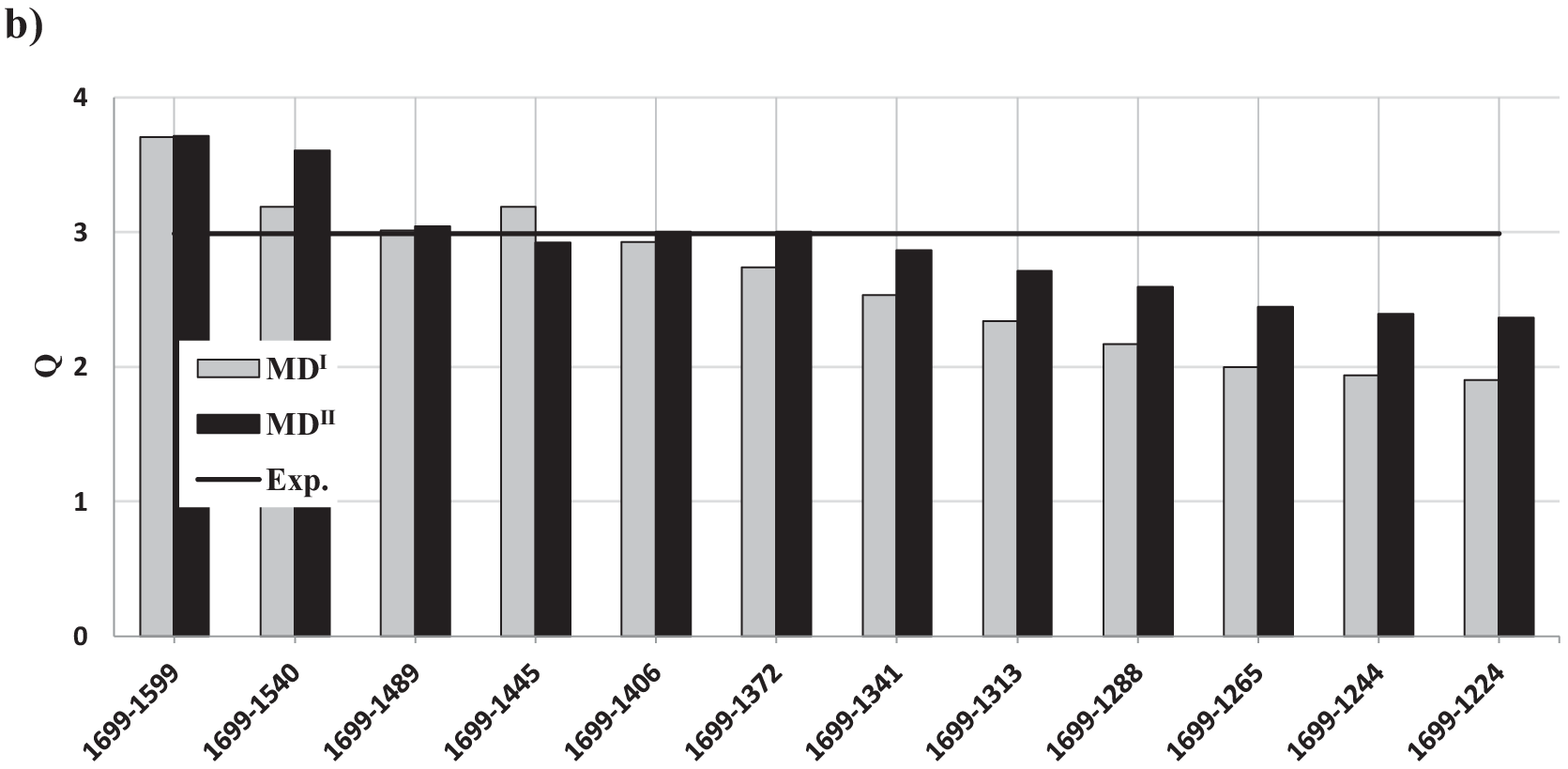}
	\end{tabular}
	\caption{Self-diffusion prefactor $D_0$ a) and activation energy $Q$ b) for different temperature intervals.}
	\label{fig:D0Q}
\end{figure}

Thereafter, using volume diffusivity results for different temperatures from our molecular dynamics simulations, Tab.\ref{tab:DiffTemp}, and the Arrhenius equation, Eq.\ref{eqn:ArrheniusLog}, Ni in B2-NiAl self-diffusion prefactor $D_0$ and activation energy $Q$ for different temperature intervals was determined, see Tab.\ref{tab:D0Q} and Fig.\ref{fig:D0Q}. The numerical results have been confronted with experimental values taken from \cite{Frank20011399}. For the temperature interval 1699-1489\,K the calculated $D_0$ and $Q$ perfectly match the experimental results for Direct(\ref{itm:first}) procedure of diffusivity determination from MD simulation results. The Extrapolation(\ref{itm:second})  procedure extends this interval into the range 1699-1372\,K. As we can see, the new approach improves agreement with experimental data, however the inconsistency problem is still visible in lower temperatures. \comm{One of the reasons explaining this issue may result from the fact that volume diffusion is major mass transport mechanism of sintering at higher temperature, frequently close to the melting point. In lower temperature, ranging from 0.5 to 0.8 of the material melting point, the grain-boundary diffusion becomes the dominant mechanism. The grain boundaries area a significant number of material defects are located, it becomes a privileged way to	atomic flow. To fully understand and estimate the diffusion parameters in lower temperatures, all types of mass transport mechanism should be analysed and take into account into considerations.}
Moreover, our NiAl diffusion $T_{simulation}$/$T_{melting}$=0.69-0.95 is much wider than that suggested by the authors in \cite{ZHOU2017331}, the low temperature results are very noisy and probably \comm{somehow} have little physical meaning for diffusion, simply the diffusion lengths are too short.

\begin{table}
	\caption{\label{tab:D0Q} Ni in B2-NiAl self-diffusion prefactor $D_0$ and activation energy $Q$ for different temperature intervals from our molecular dynamics simulations (MD$^{\text{I}}$, Direct procedure \ref{itm:first} and MD$^{\text{II}}$, Extrapolation procedure \ref{itm:second}). Experimental $D_0$=2.71-3.45\,x\,10$^{-5}$(m$^{2}$s$^{-1}$) and  $Q$=2.97-3.01(eV/atom), see \cite{Frank20011399}.}
	%	\begin{indented}
	\begin{tabular}{@{}lllll} \br
		{} & \multicolumn{2}{c}{MD$^{\text{I}}$} & \multicolumn{2}{c}{MD$^{\text{II}}$}\\ \mr
		Temp. interval[K] & {$D_0$}[m$^2$/s] & $Q$[eV/atom] & {$D_0$}[m$^2$/s] & $Q$[eV/atom]\\ \mr
		1699-1599 & 3.51317E-3 &  3.70975 & 3.90457E-3 &  3.71324 \\
		1699-1540 & 9.33994E-5 &  3.18766 & 1.82291E-3 &  3.6036 \\
		1699-1489 & 2.76108E-5 &  3.01506 & 3.40504E-5 &  3.03987\\
		1699-1445 & 9.66945E-5 &  3.19004 & 1.44477E-5 &  2.92018\\
		1699-1406 & 1.42095E-5 &  2.92579 & 2.60354E-5 &  3.00134\\
		1699-1372 & 3.61679E-6 &  2.73949 & 2.57373E-5 &  2.99977\\
		1699-1341 & 7.74201E-7 &  2.53194 & 9.39717E-6 &  2.86412\\
		1699-1313 & 1.86944E-7 &  2.34261 & 2.97039E-6 &  2.71067\\
		1699-1288 & 5.10362E-8 &  2.17135 & 1.17749E-6 &  2.58861\\
		1699-1265 & 1.38062E-8 &  2.00049 & 3.94811E-7 &  2.4458\\
		1699-1244 & 8.46639E-9 &  1.93715 & 2.5614E-7 &  2.38976\\
		1699-1224 & 6.42842E-9 &  1.90179 & 2.05872E-7 &  2.36171\\ \br
	\end{tabular}
	%	\end{indented}
\end{table}

\section{Conclusions}
\label{sec:Conc}
The volume diffusivity in the 1224\,K to 1699\,K temperature range, has been estimated in the studies reported in this paper by direct molecular statics/dynamics simulations applying the embedded atom model potential.

We can conclude that:

\begin{itemize}
	\item The diffusivity in solid state B2-NiAl intermetallic can be successfully quantified from direct MS/MD simulations.
	\item The instantaneous diffusivity stabilizes faster than the average one.
	\item The simulation times needed to achieve stabilised asymptotic diffusivity must be in the order of microseconds. 
	\item The application of the extrapolation procedure improves diffusion simulation results.
\end{itemize}

Some findings in this paper, especially relating to the simulated by MD diffusivity of stoichiometric solid state B2-NiAl intermetallic in the 1224\,K to 1699\,K temperature range and the study of required simulation time ensuring the stabilization of diffusivity results, are the first to be reported and are hopeful that will be confirmed by further research studies.
The estimated values will be used in the discrete element modelling of a sintering process of NiAl powder \cite{NosAdvPow}. 
The methodology developed in this study will be included in future multiscale sintering modelling.

%% The Acknowledgements part is started with the command \acknowledgements;
%% acknowledgements are then done as normal sections before appendix
%% \acknowledgements

%\acknowledgements
\section*{ACKNOWLEDGMENTS}
This work was supported by the National Science Centre (NCN -- Poland)  Research Project: DEC-2013/11/B/ST8/03287. Additional assistance was granted through the computing cluster GRAFEN at Biocentrum Ochota and the Interdisciplinary Centre for Mathematical and Computational Modelling of Warsaw University (ICM UW).

%\section*{References}

%% The Appendices part is started with the command \appendix;
%% appendix sections are then done as normal sections and after Acknowledgements
%% \appendix

%% \section{}
%% \label{}

%% References without bibTeX database:

%\begin{thebibliography}{-8}

%% \bibitem must have the following form:

%\small{
%\bibitem{key}

%...

%}

%\end{thebibliography}

%% References with bibTeX database:
%%\bibliographystyle{unsrt} % Bibliography style file, unsrt.bst
\bibliographystyle{elsarticle-num}
\section*{References}
\bibliography{References_arxiv}
\end{document}